\documentclass[12pt]{article}
\usepackage{graphicx,amssymb,amsfonts}
\newlength{\extraspace}
\newlength{\extraspaces}
\newcommand{\be}{\begin{equation}
\addtolength{\abovedisplayskip}{\extraspaces}
\addtolength{\belowdisplayskip}{\extraspaces}
\addtolength{\abovedisplayshortskip}{\extraspace}
\addtolength{\belowdisplayshortskip}{\extraspace}}
\newcommand{\ee}{\end{equation}}

\newcommand{\ba}{\begin{eqnarray}
\addtolength{\abovedisplayskip}{\extraspaces}
\addtolength{\belowdisplayskip}{\extraspaces}
\addtolength{\abovedisplayshortskip}{\extraspace}
\addtolength{\belowdisplayshortskip}{\extraspace}}
\newcommand{\ea}{\end{eqnarray}}

\newcommand{\nonu}{\nonumber \\[.5mm]}
\newcommand{\A}{&\!\!\!}

\newcommand{\newsection}[1]{
\vspace{7mm} \pagebreak[3] \addtocounter{section}{1}
\setcounter{subsection}{0} \setcounter{footnote}{0}
\begin{center}
{\large {\bf \thesection. #1}}
\end{center}
\nopagebreak
\medskip
\nopagebreak \hspace{3mm}}

\begin{document}

\begin{center}
{\bf   Schwarzschild solution in extended teleparallel gravity}\footnote{ PACES numbers: 04.50. Kd, 04.70.Bw, 04.20. Jb\\
\hspace*{.5cm}
 Keywords: f(T) theories of gravity, exact solution,
energy}
\end{center}
\begin{center}
{\bf Gamal G.L. Nashed}
\end{center}

\bigskip

\centerline{\it Centre  for Theoretical Physics, The British
University in  Egypt} \centerline{\it Sherouk City 11837, P.O. Box
43, Egypt \footnote{ Mathematics Department, Faculty of Science, Ain
Shams University, Cairo, 11566, Egypt \\
\hspace*{.2cm} Egyptian Relativity Group (ERG) URL:
http://www.erg.eg.net}}

\bigskip
\centerline{ e-mail:nashed@bue.edu.eg}

\hspace{2cm} \hspace{2cm}
\\
\\
\\
\\
\\

 Tetrad field, with  two unknown functions of radial coordinate and an angle $\Phi$ which is the polar angle $\phi$ times a function of the redial coordinate, is applied to the field equation of  modified theory of gravity. Exact  vacuum solution  is derived whose scalar torsion, $T ={T^\alpha}_{\mu \nu}
{S_\alpha}^{\mu \nu}$, is constant. When the angle $\Phi$ coincides with the polar angle $\phi$, the derived solution will be a solution only for linear form of $f(T)$ gravitational theory.
\begin{center}
\newsection{\bf Introduction}
\end{center}
Common consensus in the scientific community is that the characterization of the gravitational field powered by Einstein general relativity (GR) theory. This theory is bound to miss at scales of the magnitude of the Planck length, in which the space-time frame should be clarified by a quantum regime. Furthermore, ultimate of the physical phenomena, GR also faces a curiosity problem related to the late cosmic speed up stage of the Universe. Due to the previous problems and for other defects, i.e. dark energy, dark matter etc., GR has been the topic of many modifications. These modifications have been attempted  to supply a most satisfying qualification  of the gravitational field in the above aforementioned extreme regimes. One of the most modified gravitational theories is the $f(T)$ gravity. This theory constructed in a space-time having absolute parallelism \cite{HHK}-\cite{HS9}. In this space-time, the curvature contributions vanishing identically and the only  contribution  is due to the anti symmetric part of the non-symmetric affine connection. This procedure is of the so-called Weitzenb\"ok$'$s, Teleparallel Equivalent of General Relativity (TEGR), space-time

Recently, $f(T)$ gravity theory has been elaborated in specifics. Many of $f(T)$ gravity theories had been analyzed in \cite{BF9}-\cite{WY1}. It is
found that $f(T)$ gravity theory is not synonymous dynamically to TEGR  Lagrangian through  conformal
transformation \cite{Yr1}. Many observational restrictions  had been studied  \cite{Bg1}-\cite{WMQ}. Large-scale structure in $f(T)$ gravity theory
had been analyzed \cite{LSB}. Perturbations in the area of cosmology in $f(T)$ gravity had been demonstrated  \cite{DDS}-\cite{CCDDS}. Birkhoff's theorem,
 in $f(T)$ gravity had  been studied \cite{MW1}. Stationary solutions having spherical symmetry have
been derived for  $f(T)$ theories \cite{Wt1}. Relativistic Stars and the cosmic expansion
have been  studied \cite{Dy}.

$f(T)$ gravitational theories have been engaged many concerns and it had been indicated  that the Lagrangian  and the  equations of motion  of those theories are not variant under local Lorentz transformations \cite{LSB1}. It had been explained that the reasons why setting back local Lorentz symmetry in $f(T)$ theories cannot upgrade  to credible dynamics, even  if one relinquishes teleparallelism \cite{SLB1}. The equations of motion of $f(T)$  theories have been stated to be  differ from those of $f(R)$ theories \cite{NO}$-$\cite{BMT}, because they are of second order instead of  fourth order. Such property has been believed   as an indicator, which shows that the theory might be of much interest than this of GR. Because of the non-locality of these theories, $f(T)$, it seems  to contain more degrees of freedom.

The aim of the present study is to find analytic vacuum spherically symmetric solution, in the framework $f(T)$ gravitational  theory.

 In \S 2, a brief review of the $f(T)$ gravitational theory is provided.  Also in \S 2, non-diagonal, spherically symmetric tetrad field with two
 unknown functions of  radial coordinate in addition to the angle $\Phi$ is
given. Application  of such tetrad   to the field equation of $f(T)$ is
provided.  Analytic vacuum spherically symmetric
solution with one constant of integration is derived in \S 2. In \S 3, the physical properties of the derived solution, i.e.,  the decomposition
of the derived solution is achieved and
 the energy  is calculated to understand the physical meaning of the constant of integration. Final section is devoted to discussions.
\newpage
\newsection{Brief review of f(T) gravitational theory and spherically symmetric solution}

The equation of motions of $f(T)$ gravitational theory have the form \cite{BF9} \be {S_\mu}^{\nu \rho}
T_{,\rho} \
f(T)_{TT}+\left[e^{-1}{e^a}_\mu\partial_\rho\left(e{e_a}^\alpha
{S_\alpha}^{\nu \rho }\right)+{T^\alpha}_{\lambda
\mu}{S_\alpha}^{\nu
\lambda}\right]f(T)_T+\frac{1}{4}\delta^\nu_\mu f(T)=4\pi {{\cal
T}^\nu}_\mu,\ee where ${T^\alpha}_{\mu \nu}  \stackrel {\rm def.}{=}
{\Gamma^\alpha}_{\nu \mu}-{\Gamma^\alpha}_{\mu \nu} ={e_a}^\alpha
\left(\partial_\mu{e^a}_\nu-\partial_\nu{e^a}_\mu\right)$ is the torsion tensor, ${\Gamma^\alpha}_{\mu \nu}$ is the non-symmetric affine connection, ${e_a}^\alpha$ is the tetrad field which is the main block in the modified theories of teleparallel  of gravity and $e = det ({e^\mu}_a) =\sqrt{-g}$, is the determinant of the tetrad. the tensor ${S_\alpha}^{\nu
\lambda}$ is defined as follows ${S_\alpha}^{\mu \nu}
\stackrel {\rm def.}{=} \frac{1}{2}\left({K^{\mu
\nu}}_\alpha+\delta^\mu_\alpha{T^{\beta
\nu}}_\beta-\delta^\nu_\alpha{T^{\beta \mu}}_\beta\right)$ where ${K^{\mu \nu}}_\alpha  \stackrel {\rm def.}{=}
-\frac{1}{2}\left({T^{\mu \nu}}_\alpha-{T^{\nu
\mu}}_\alpha-{T_\alpha}^{\mu \nu}\right)$ is the contortion and $T$ is the scalar torsion which is defined as $ T \stackrel {\rm def.}{=} {T^\alpha}_{\mu \nu}
{S_\alpha}^{\mu \nu}$. $T_{,\rho}=\frac{\partial T}{\partial
x^\rho}$, $f(T)_T=\frac{\partial f(T)}{\partial T}$,
$f(T)_{TT}=\frac{\partial^2 f(T)}{\partial T^2}$ and ${{\cal
T}^\nu}_\mu$ is the energy momentum tensor.

The total energy-momentum of $f(T)$ gravitational theory contained
in a three-dimensional volume $V$ has the form \cite{US3}  \begin{equation} P^a=\int_V d^3x
\ e \ {e^a}_\mu \ t^{0 \mu}=\frac{1}{4\pi}\int_V d^3x  \partial_\nu\left[e{S}^{a 0
\nu} f(T)_T\right].\end{equation}  In this study we are
interested in studying the  vacuum case of $f(T)$  theory,
i.e., ${\cal T}^\nu_\mu=0$.

Assuming that the space-time possessing a stationary and spherical
symmetry  has the form
\be \left( {e^\mu}_a \right)= \left( \matrix{A(r) & 0& 0 &
0\vspace{3mm} \cr 0 &B(r)\sin\theta
\cos\Phi&r\cos\theta \cos\Phi & -r\sin\theta \sin\Phi \vspace{3mm}
\cr 0&B(r)\sin\theta \sin\Phi&r\cos\theta
\sin\Phi & r\sin\theta \cos\Phi \vspace{3mm} \cr 0 &
B(r)\cos\theta&-r\sin\theta & 0 \cr } \right),\ee
 where $A(r)$ and $B(r)$ are two unknown functions of the radial coordinate, $r$ and $\Phi=\phi L(r)$.

    Using Eq. (1) and
  $ \left({e^i}_{  \mu} \right)$ given by Eq. (3), one can obtain $e = det ({e^\mu}_a) = r^2 AB\sin \theta$\footnote{For briefing we will denote $A\equiv A(r)$, $B\equiv B(r)$ and $L\equiv \phi L(r)$.}
   and the
torsion scalar and its derivatives  in the form  \ba
\A \A T(r)=-\frac{2\left( A B^2L_\phi-AB-rBA'-AB L_{\phi}+A+2rA'-rBA'L_{\phi}\right)}{r^2AB^2}, \nonu
\A \A \nonu
\A \A   \textrm{where} \qquad L_{\phi}=\frac{\partial L(\phi)}{\partial \phi}, \qquad \textrm{and} \qquad A'=\frac{\partial A(r)}{\partial r},
\qquad B'=\frac{\partial B(r)}{\partial r}, \nonu
\A \A \nonu
\A \A T'=\frac{\partial T}{\partial r}=-\frac{1}{r^3A^2B^3}\Biggl(2r^2B[AA''-A'^2][B-2+B L_\phi]-2rAA'\Biggl\{BL_\phi[B+rB']+rB'[B-4]\nonu
\A \A-2B+B^2\Biggr\}-2A^2\Biggl\{BL_\phi[2B-2B^2+rB']+rB'[B-2]+2B^2-2B\Biggr\}\nonu
\A \A T_\phi=\frac{\partial T}{\partial \phi}=\frac{2L_{\phi \phi}[rA'-A[B-1]]}{r^2AB}, \qquad \qquad L_{\phi \phi}=\frac{\partial^2 L(\phi)}{\partial \phi^2}.\ea Using the above calculations, the field equations (1)  take the form \ba \A \A
4\pi{\cal T}_0^0=-\frac{f_{TT}T'[B-2+B L_\phi]}{r^4A^2B^5} +\frac{f_{T}AB^2r^2}{2r^4A^2B^5}\Biggl(B^2L_\phi\{rA'-A(B-1)\}\nonu
\A \A +rBA'(B-2)+A(2rB'-2B+B^3+B^2)\Biggr)+\frac{f}{4}\,\ea
\be 4\pi{\cal T}_1^1=\frac{f_{T}
\{2AB^2+2rBA'L_\phi-8rA'-4A+2AB-2AB^2L_\phi+2BrA'+2ABL_\phi
\}}{4r^2AB^2}+\frac{f}{4}\,\ee \be 4\pi{\cal
T}_1^2=-\frac{f_{TT}T'\cot\theta[L_\phi-1]}{2r^5A^2B^3},\ee\ba \A \A 4\pi{\cal
T}_2^2=-\frac{f_{TT}T'[rA'-ABL_\phi+A]}{r^5A^2B^3}-\frac{f_{T}}{2r^2AB^3}\Biggl(r^2BA''-rA'[B^2-3B+rB'+B^2L_\phi]\nonu
\A \A-A[B^2(1-B)L_\phi+rB'+B^2-B]\Biggr)+\frac{f}{4},\ea
\be 4\pi{\cal T}_3^1=\frac{f_{TT}
\{L_{\phi \phi}[AB-A-rA']^2
\}}{r^3A^2B^3}\,\ee
\ba \A \A 4\pi{\cal
T}_3^3=-\frac{f_{TT}T'[rA'-A(B-1)]}{2r^4A^3B^5}-\frac{f_{T}}{2r^2AB^3}\Biggl(r^2BA''-rA'[B^2-3B+rB'+B^2L_\phi]\nonu
\A \A-A[B^2(1-B)L_\phi+rB'+B^2-B]\Biggr)+\frac{f}{4}.\ea

From equations (5)--(10), it is clear that $A\neq 0$ and $B\neq 0$. To solve the differential equations (5)-(10) we put the
following constrains  \ba \A \A T=const.=T_0\Rightarrow T'=0, \qquad \qquad L_{\phi \phi}=0\Rightarrow T_\phi=0,\nonu
\A \A B^2L_\phi\{rA'-A(B-1)\}+rBA'(B-2)+A(2rB'-2B+B^3+B^2)=T_0,\nonu
\A \A 2AB^2+2rBA'L_\phi-8rA'-4A+2AB-2AB^2L_\phi+2BrA'+2ABL_\phi=T_0,\nonu
\A \A r^2BA''-rA'[B^2-3B+rB'+B^2L_\phi]-A[B^2(1-B)L_\phi+rB'+B^2-B]=T_0.\ea The first constraint of Eq. (11) ensures the vanishing of the right hand side of
${\cal T}_1^2$ and ${\cal T}_3^1$ and also the disappears of $f_{TT}$ in Eqs. (5), (8) and (10). The rest of constraints of Eq. (11) constitute three
non-linear differential equations in three unknown functions, $A(r)$, $B(r)$ and $L(\phi)$. The solution of these differential equations has the form

  \be A=\frac{1}{B}=\sqrt{1-\frac{c_1}{r}}, \qquad L(\phi)=-\frac{\left\{2\sqrt{r-2c_1}+\sqrt{r}(2+r^2T_0)\right\}\phi}{2\sqrt{r-2c_1}+2\sqrt{r}},\ee where $c_1$ is a constant of integration. Using Eq. (12)  in
Eq. (4) we get a constant value of the scalar torsion which
gives a vanishing quantity of the second equation of Eq. (4). Therefor, Eq. (12) is an exact vacuum solution to equations (5)--(10) provided that \be f(T_0)=-T_0,
 \qquad f_T(T_0)=1, \qquad f_{TT}\neq0.\ee
 To understand the nature of the constant appears in Eq. (12) we are going to discuss the physics related to this solution and
 calculating the energy associated with the
tetrad  field (3) after using Eq.  (12).
\newsection{Physical properties of the derived solution}

To understand the construction of the derived solution let us
rewrite tetrad (3) after using solution (12) in the following form
 \ba \A \A \left( {e^i}_\mu \right)= \left({\Lambda^i}_j\right) \left( {e^j}_\mu \right)_d \qquad \textrm {where} \nonu
 \A \A \nonu
\A \A \left({\Lambda^i}_j\right)=\left( \matrix{ 1 &  0 & 0 & 0
\vspace{3mm} \cr  0  &  \sin\theta \cos\Phi &  \cos\theta \cos\Phi &
- \sin\Phi \vspace{3mm} \cr 0  & \sin \theta \sin \Phi& \cos\theta
\sin\Phi & \cos\Phi \vspace{3mm} \cr 0  & \cos\theta & -\sin\theta &
0 \cr }\right)\;, \qquad    \left( {e^j}_\mu \right)_d=\left( \matrix{
\sqrt{1-\frac{c_1}{r}} & 0 & 0 & 0 \vspace{3mm} \cr 0 &
\displaystyle\frac{1}{\sqrt{1-\frac{c_1}{r}}} &0& 0 \vspace{3mm} \cr
0  & 0&r &0 \vspace{3mm} \cr 0  & 0 & 0 & r\sin\theta \cr
}\right)\;.\nonu
\A \A \ea
Eq. (14) shows that the tetrad (3) with solution (12) consists of a diagonal tetrad in addition to ``so(3)''. The diagonal tetrad alone is not a solution
to the field equations of $f(T)$. Therefore, ``so(3)'' plays an important role in $f(T)$  with the angle $\Phi$ which is
a function of the radial coordinate $r$. It is of interest to note that $\Phi=\phi L(r)=\phi$, the derived solution will be a solution only  to the linearized form of $f(T)$, i.e., $f(T)=T$.

Now we are going to calculate the energy associated with the derived solution using formula (2).
The necessary  non-vanishing
components of the tensor $S^{\mu \nu \rho}$ are  \begin{eqnarray} & & S^{001}=-\frac{r+2\sqrt{r^2-2rc_1}+L_\phi}{\sqrt{r^4-2^3rc_1}}, \nonu
& & {\textrm where} \qquad L_\phi=\frac{\partial L(\phi)}{\partial \phi}, \qquad E=P^0=\frac{1}{4\pi}\int_V d^3x  \partial_\nu\left[e
{e^a}_\mu{S}^{\mu 0 \nu} f_T\right], \qquad f_T\approx1+T+T^2+\cdots,\nonumber\\
& &  E\approx\left(M-r-\frac{rM^2 T_0}{2}-2T_0r^3\right)\Biggl[1+T_0+{T_0}^2+\cdots\Biggr]\approx\left(M-r-\frac{rM^2 T_0}{2}-2T_0r^3\right),\end{eqnarray}
which is divergence. To remove such divergent we use the following
expression:
\begin{equation} {P^a}_{\textrm{Regularized}}\approx \frac{1}{4\pi}\int_V d^3x \left( \left\{ \partial_\nu\left[e{S}^{a 0
\nu} f(T)_T\right]\right\}-\left\{ \partial_\nu\left[e{S}^{a 0 \nu}
f(T)_T\right]\right\}_{\textrm {vanishing \  physical \
quantity}}\right),\end{equation} where the physical quantity here is $c_1$. Using equation (15) we get:
\begin{equation}
E={P^0}_{\textrm{Regularized}}\approx c_1,\end{equation} which is the energy of Schwarzschild  provided that $c_1=M$ where $M$ is the gravitational mass \cite{Nprd}.

\newsection{Main results and discussion}

   In this study we have considered the modified gravitational theory,  $f(T)$, in the vacuum case. The field equations   have been applied to a non-diagonal   tetrad field having two unknown functions in the redial coordinate and an angle $\Phi$ which is a function of the radial coordinate $r$. Six non-linear differential equations have been derived. Some constraints have been applied to solve these differential equations. These constrains constitute three non-linear differential equations in three unknown functions. The solution of these constrains contains one constant of integration.   Therefore, exact vacuum spherically symmetric solution to the field equations of $f(T)$ gravitational theory has been derived. This solution has a constant scalar torsion, i.e. $T=T_0$ and satisfies the field equations of $f(T)$  if  Eq. (13) is satisfied.  To understand what the nature of the constant of integration is we calculate the energy associated with the derived solution. We have   shown that such constant is related to the mass of gravitation.

We have shown that the tetrad of the derived solution can be rewritten as two matrices. The first matrix is ``so(3)'' which is a special case of Euler's angle \cite{Nprd3}. On the other hand, the second matrix is a diagonal matrix of Schwarzschild metric space-time. We have shown that when $L(\phi)=\phi$ the derived solution will be a solution to the first order of $f(T)$.

If  one repeated the same procedure done in the derivation of the solution, we can derive for tetrad (3) another solution in the framework of $f(T)$. This solution has the following form

\be A=\frac{1}{B}=\sqrt{1-\frac{c_2}{r}}, \qquad L(\phi)=-\phi,\ee where $c_2$ is another constant related to the gravitational mass. Solution (18) gave a vanishing value of the scalar torsion, $T=0$,  and is a solution to the field equations of $f(T)$ provided that
\be f(0)=0,
 \qquad f_T(0)\neq 0, \qquad f_{TT}\neq0.\ee

\end{document}